\documentclass[%
reprint,
unsortedaddress,
showpacs,preprintnumbers,
 amsmath,amssymb,
 aps,
pre,
floatfix
]{revtex4-1}

\usepackage{dcolumn}
\usepackage{bm}
\usepackage{color}
\usepackage{comment}
\usepackage{graphicx}
\usepackage{CJK}

\allowdisplaybreaks[4]

\newcommand{\ustb}{%
Department of Applied Mathematics, University of Science and Technology Beijing,
Beijing 100083, China}

\newcommand{\camtsinghua}{%
Zhou Pei-Yuan Center for Applied Mathematics,
Tsinghua University, Beijing 100084, China}

\begin{document}

\title{Relaxation-rate formula for the entropic lattice Boltzmann method}%

\author{Weifeng Zhao}%
\email[Email: ]{wfzhao@ustb.edu.cn}
\affiliation{\ustb}

\author{Wen-An Yong}%
\email[Email: ]{wayong@tsinghua.edu.cn}
\affiliation{\camtsinghua}

\date{\today}

\begin{abstract}

An elegant and uniform relaxation-rate formula is presented for the entropic lattice Boltzmann method (ELBM). The formula not only guarantees the discrete time H-theorem at numerical level but also gives full consideration to the consistency with hydrodynamics. With this novel formula, the computational cost of the ELBM is significantly reduced and the method now can be efficiently used for a broad range of hydrodynamics applications including high Renolds number flows. Moreover, we demonstrate that the grid points where flow fields change drastically are effectively marked by the formula.

\end{abstract}



\maketitle



In the last three decades, the lattice Boltzmann method (LBM) has become a popular mesoscopic numerical method for fluid dynamics, with applications ranging from high Reynolds number flows to flows at porous media and relativistic hydrodynamics \cite{Su,Gu}. We refer to \cite{BSV,CD,AC} for reviews of the method and its applications.
As a discrete space-time kinetic theory for hydrodynamics, the LBM employs discretized particle distribution functions associated with discrete velocities to describe the flow field \cite{Su,Gu}. By fitting the discrete velocities into a regular lattice, the LBM realizes the propagation and collision of distribution functions efficiently.

Though simple and efficient, the standard LBM is limited to moderate Reynolds number flows due to the lack of numerical stability \cite{Su,AK,BYCW}. To alleviate this obstacle, the entropic LBM (ELBM) restoring the discrete H-theorem has been proposed in \cite{KGSB,KFO,AK2002,AK2002jsp,AKO2003,BLY2004,CAK2006,KVYSV2007,KBC2014,AKTA2017} as a paradigm shift for computational fluid dynamics. The introduction of the discrete H-theorem in the ELBM significantly extends the operation range of the discrete kinetic theory to turbulent flows \cite{CFKTB2010,DBCBK2016}, binary droplet collisions \cite{MCK2015,MCK2016}, multiphase fluid-solid interface problems \cite{MCK2015pre} and compressible flows \cite{FCK2015}.
The solid physical background and successful applications make the ELBM a powerful approach for the study of complex flows.

A key step in the ELBM is to determine the relaxation-rate involving a parameter $\alpha$ introduced in \cite{KFO} for ensuring the discrete H-theorem. It needs to solve a complicated nonlinear algebraic equation, which greatly affects the efficiency of the ELBM. Though considerable efforts have been made in the past \cite{AK2002, AK2002jsp,CAK2006,TUSK2007} to improve the efficiency, it was only recognized by Brownlee {\it et al.}~\cite{BGL2007} that the computational value of $\alpha$ should not lead to a numerical entropy increase. This requirement essentially guarantees the discrete H-theorem at the numerical level and is further emphasised in \cite{AKTA2017} recently.  Therein some analytical approximate expressions of $\alpha$, which also guarantee the discrete H-theorem numerically, are derived by relaxing the entropy equality \cite{AKTA2017} and by making a near-equilibrium assumption. Nevertheless, the first-order approximation is too dissipative while higher-order approximations are difficult to be explicitly obtained. Therefore, a critical breakthrough is much needed for the efficient determination of $\alpha$.

In this Letter, we solve this long-standing problem by proposing an elegant and uniform formula for the parameter $\alpha$,  or equivalently the relaxation-rate of the ELBM. This formula is based on a novel combination of the consistency of the ELBM and the constraint that the entropy must not increase within a discrete time step. Besides compliance of the discrete H-theorem at numerical level, an excellent property of the formula is that it is applicable to arbitrary convex entropy functions. Additionally, numerical simulations demonstrate that the grid points where flow fields change drastically are effectively marked by the formula.

Before presenting our formula, we recall from \cite{AKO2003} that the entropic lattice Boltzmann method (ELBM) reads as
{\small
\begin{equation}\label{1}
  f_i(\bm{x} + \bm{c}_i \delta_t, \, t+ \delta_t )
  =
  f_i(\bm{x},\, t) + \alpha \beta (f_i^{(eq)}(\bm{x},\, t) - f_i(\bm{x},\, t) )
\end{equation}
}
with $i=1,2,\cdots, N$. Here $f_i=f_i(\bm x, t)$ is the $i$-th distribution function for particles with velocity $\bm c_i$ at position $\bm x$ and time $t$,
$\delta_t$ is the time step, $\alpha \beta$ is the relaxation-rate with $\beta \in (0,1)$ related to the fluid viscosity $\nu$ via
\begin{equation}\label{2}
\beta = \frac{ \delta_t c_s^2}{ 2 \nu + \delta_t c_s^2 },
\end{equation}
$c_s$ is the sound speed and $f_i^{(eq)}=f_i^{(eq)}(\bm{x},\, t)>0$ is the equilibrium minimizing the convex entropy function
\begin{equation} \label{3}
H = H(f)=\sum_i f_i \ln( f_i / W_i )
\end{equation}
subject to the conservation laws of mass and momentum (for the isothermal case):
\begin{equation*}
\sum_i f_i^{(eq)} = \rho \equiv \sum_i f_i, \quad  \sum_i \bm c_i  f_i^{(eq)} = \rho \bm u \equiv \sum_i \bm c_i  f_i.
\end{equation*}
In the above equations, $f$ stands for the vector $(f_1, f_2, \cdots, f_N)$, $W_i>0$ is the $i$-th weight and $\rho$ and $\bm u$ are the macroscopic fluid density and velocity, respectively.
A key point of the ELBM is the parameter $\alpha$ in \eqref{1} that maintains the entropy balance
\begin{equation}\label{4}
H(  f + \alpha( f^{(eq)} - f )) = H(f).
\end{equation}
If $\alpha=2$ and $f_i^{(eq)}$ is taken as polynomials, \eqref{1} degenerates to the standard lattice Bhatnagar-Gross-Krook (LBGK) model \cite{QDL}.

With the equal entropy
auxiliary distribution $f^{*}:=f + \alpha( f^{(eq)} - f )$, the ELBM \eqref{1} can be rewritten as
\begin{subequations}\label{5}
\begin{align}
& {\tilde f}_i(\bm{x} , \, t) =
  (1-\beta)f_i(\bm{x},\, t)  +  \beta f_i^{*}(\bm{x},\, t) , \label{5a} \\
& f_i(\bm{x} + \bm{c}_i \delta_t, \, t+ \delta_t ) = {\tilde f}_i(\bm{x} , \, t). \label{5b}
\end{align}
\end{subequations}
By the convexity of the function $H$ defined in \eqref{3}, we see from \eqref{4} and \eqref{5a} that
{\small
\begin{equation*}
  H({\tilde f}(\bm{x} , \, t))  \leq (1-\beta) H(f(\bm{x},\, t)) +
   \beta H(f^{*}(\bm{x},\, t))  = H(f(\bm{x},\, t)).
\end{equation*}
}
Furthermore, for a periodic domain $\Omega$ we have
{\small
\begin{equation}\label{6}
  \sum_{\bm x \in \Omega} H(f(\bm x, t+\delta_t))  = \sum_{\bm x \in \Omega}H({\tilde f}(\bm{x} , \, t)) \leq \sum_{\bm x\in \Omega}H(f(\bm{x},\, t)).
\end{equation}
}
This is a discrete H-theorem for the ELBM.

In spite of this H-theorem, the entropy balance equation \eqref{4} requires an additional step of searching for the parameter $\alpha$. The efficiency of this search is crucial for the realization of the ELBM.

Here we derive a simple approximate solution formula for $\alpha$.
Observe that the discrete H-theorem \eqref{6} holds true for all $f^{*}(\alpha):=f + \alpha (f^{(eq)}-f)$ satisfying
\begin{equation}\label{7}
H( f^{*} (\alpha)) \leq H(f),
\end{equation}
instead of the entropy balance \eqref{4}.
Then we can relax Eq.~\eqref{4} and replace it with the inequality \eqref{7} as in \cite{AKTA2017}. Based on this observation, we propose an efficient implementation different from that in \cite{AKTA2017}.
In what follows, we assume that $f^{(eq)}\ne f$. Otherwise, $\alpha$ can be any number.

\begin{figure}[!ht]
\begin{center}
\includegraphics[scale=0.37]{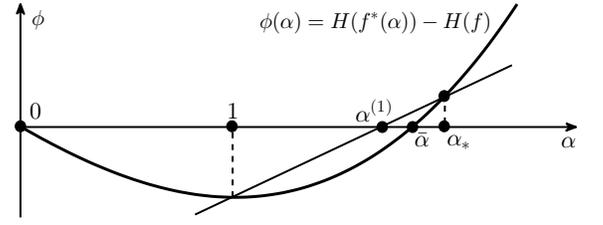}
\caption{Graph of the function $H(f^*(\alpha)) - H(f)$ which is convex and increasing on $[1, \alpha_{\max}]$.  $\bar{\alpha}$ is the solution to Eq.~\eqref{4}.}
\label{Fig:secant_1iter}
\end{center}
\end{figure}

First, we follow \cite{AK2002jsp} and define for non-negative $f_i$:
\begin{equation*}
\alpha_{\max} = \min_{i: f_i^{(eq)} < f_i } \left\{  \frac{f_i}{f_i - f_i^{(eq)}} \right\}>1,
\end{equation*}
which partly measures the departure of distribution $f$ from equilibrium.
Because $\sum_if_i = \sum_if_i^{(eq)}$, the above set is non-empty. Notice that $\alpha\in[0, \alpha_{\max}]$ ensures the nonnegativity of the distribution $f^{*}(\alpha)$ and thereby the entropy function $H(f^{*}(\alpha))$ is well defined \cite{AK2002jsp}. Moreover, from the convexity of $H(f)$ it follows that $H(f^{*}(\alpha))$ is convex and increasing on $[1, \alpha_{\max}]$ (see Fig.~\ref{Fig:secant_1iter} and the Supplementary Material).

Recall from \cite{AK2002jsp} that the viscosity relation \eqref{2} is derived with $\alpha=2$. Then $\alpha$ should be close to 2 as much as possible in order to maintain the consistency. To this end, we introduce
\begin{equation}\label{8}
\alpha_*=\min\{2, \alpha_{\max}\}.
\end{equation}
Notice that $\alpha_{\max}\gg 1$ near equilibrium.
If $H(f^{*}(\alpha_*)) \leq H(f)$, then $\alpha$ is taken as $\alpha_*$, which maintains the entropy inequality \eqref{7}. Otherwise, we refer to Fig.~~\ref{Fig:secant_1iter} and take

\begin{align}\label{9}
\alpha = \alpha^{(1)}
:=  & \alpha_*
+ \frac{ H( f^{*}(\alpha_*)) - H(f) }{ H( f^{*}(\alpha_*)) - H(f^{(eq)})  } ( 1 - \alpha_*  ).
\end{align}
From Fig.~\ref{Fig:secant_1iter} we can see that this $\alpha^{(1)}$ will be very close to the solution of the entropy balance equation \eqref{4} if $[H( f^{*}(\alpha_*)) - H(f)]$ is small, which is of frequent occurrence.
Thanks to the convexity of function $H$, it is proved in the Supplementary Material that
$$
H( f^{*}(\alpha^{(1)} )) \leq H(f).
$$

\begin{figure}[!ht]
\begin{center}
\includegraphics[scale=0.35]{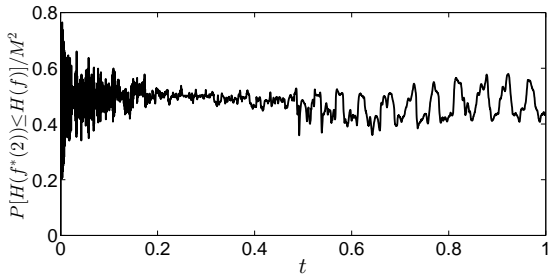}
\caption{The proportion $P[H(f^{*}(2)) \leq H(f)]/M^2$ as a function of time $t$, where $P[H(f^{*}(2)) \leq H(f)]$ denotes the number of grid points where $H(f^{*}(2)) \leq H(f)$ and $M^2$ is the total grid point number.}
\label{Fig:ratio}
\end{center}
\end{figure}

About the above implementation, three remarks are in order. (a). The implementation not only guarantees the nonnegativity of distributions but also maintains the discrete H-theorem \eqref{6}. The former is due to $\alpha\in[0, \alpha_{\max}]$ and the latter follows from $H(f^{*}(\alpha_*)) \leq H(f)$ or $H(f^{*}(\alpha^{(1)})) \leq H(f)$.
(b). The introduction of $\alpha_*(\leq 2)$ is a key in our implementation, it reduces the computational cost drastically. Indeed, $\alpha_*$ is often much smaller than $\alpha_{\max}$ used in \cite{AK2002jsp,AK2002,TUSK2007,BGL2007}.
It also extracts lots of grid points where the entropy balance \eqref{4} is irrelevant. Moreover, our numerical example shows that the number of grid points where $H(f^{*}(2)) \leq H(f)$ is about half of the total grid point number, see Fig.~\ref{Fig:ratio}.
(c). Formula \eqref{9} is much simpler than those given in \cite{AKTA2017}. It relies neither on any near-equilibrium assumption nor on the specific form of the entropy function $H=H(f)$, while those in \cite{AKTA2017} do.

\begin{figure}[!ht]
\begin{center}
\includegraphics[scale=0.3]{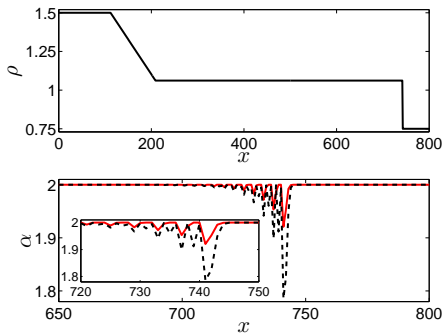}
\caption{Top: The exact density profile for the shock tube problem at time $t=500$.
The initial state is $\rho(0\leq x \leq 400) = 1.5$, $\rho(400 < x \leq 800) = 0.75$ and $u(0 \leq x \leq 800) = 0$.
Bottom: Distributions of $\alpha$ with viscosity $\nu = 10^{-5}$ at the compressive shock front from our implementation (red solid line) and the first-order EELBM (black dashed line).}
\label{Fig:1D_alpha}
\end{center}
\end{figure}

To compare Formula \eqref{9} and the essentially ELBM (EELBM) with its first-order approximation \cite{AKTA2017}, we simulate the one-dimensional shock tube problem in \cite{AK}. The exact density profile is displayed in Fig.~\ref{Fig:1D_alpha} and the two implementations both produce solutions oscillating near the shock. We see that
in a narrow region of the shock front, $\alpha$ obtained with the implementations is not larger than 2 and our implementation gives larger $\alpha$ than the EELBM. At the point of maximum departure, the deviation of $\alpha$ from 2 is 10.71\% for the EELBM while that for the present implementation is only 3.96\%. This clearly shows that Formula \eqref{9} gives a better approximation to the solution of the equation \eqref{4} than the first-order EELBM.

Furthermore, we simulate the double shear flow with periodic boundary conditions. For this problem, the spatial domain is $[0,1] \times [0,1]$, and the initial state is \cite{MB1997,KBC2014}
\begin{equation*}
\begin{aligned}
& u_x(x, y, 0) = \left\{
\begin{aligned}
& U \tanh[ \lambda( y - 0.25 ) ], \quad y \leq 1/2, \\
& U \tanh[ \lambda( 0.75 - y ) ], \quad y > 1/2,
\end{aligned}
\right. \\
& u_y(x,y,0) = 2 \times 10^{-3} \sin[ 2\pi(x+0.25) ] \\
\end{aligned}
\end{equation*}
and $\rho(x,y,0) = 1$. Here $u_x$ and $u_y$ are the $x-$ and $y-$component of the fluid velocity, respectively, $U=0.04$ and $\lambda=80$ determines the slop of the shear layer.

\begin{figure*}[!ht]
\begin{center}
\includegraphics[scale=0.25]{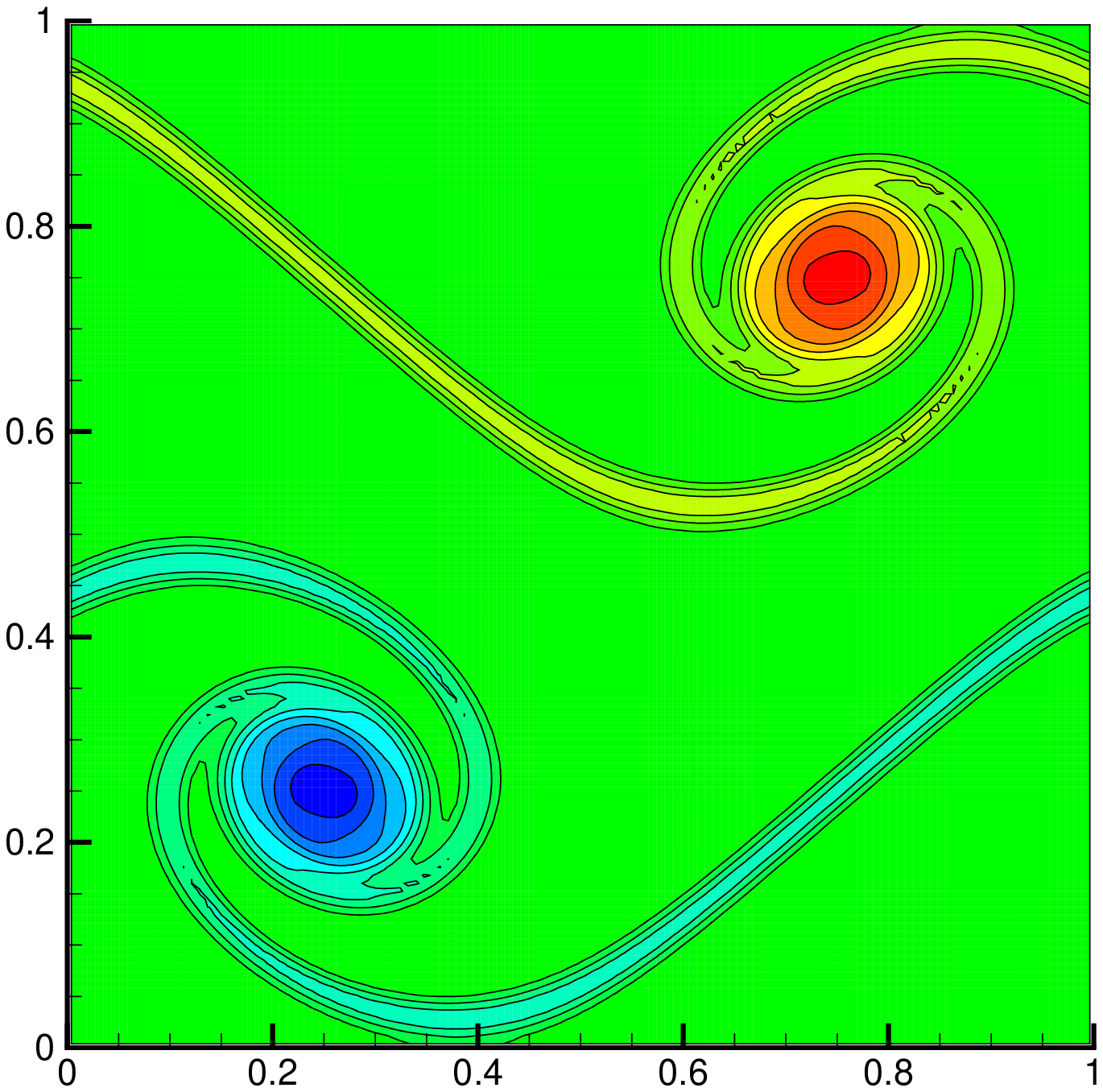}
\includegraphics[scale=0.25]{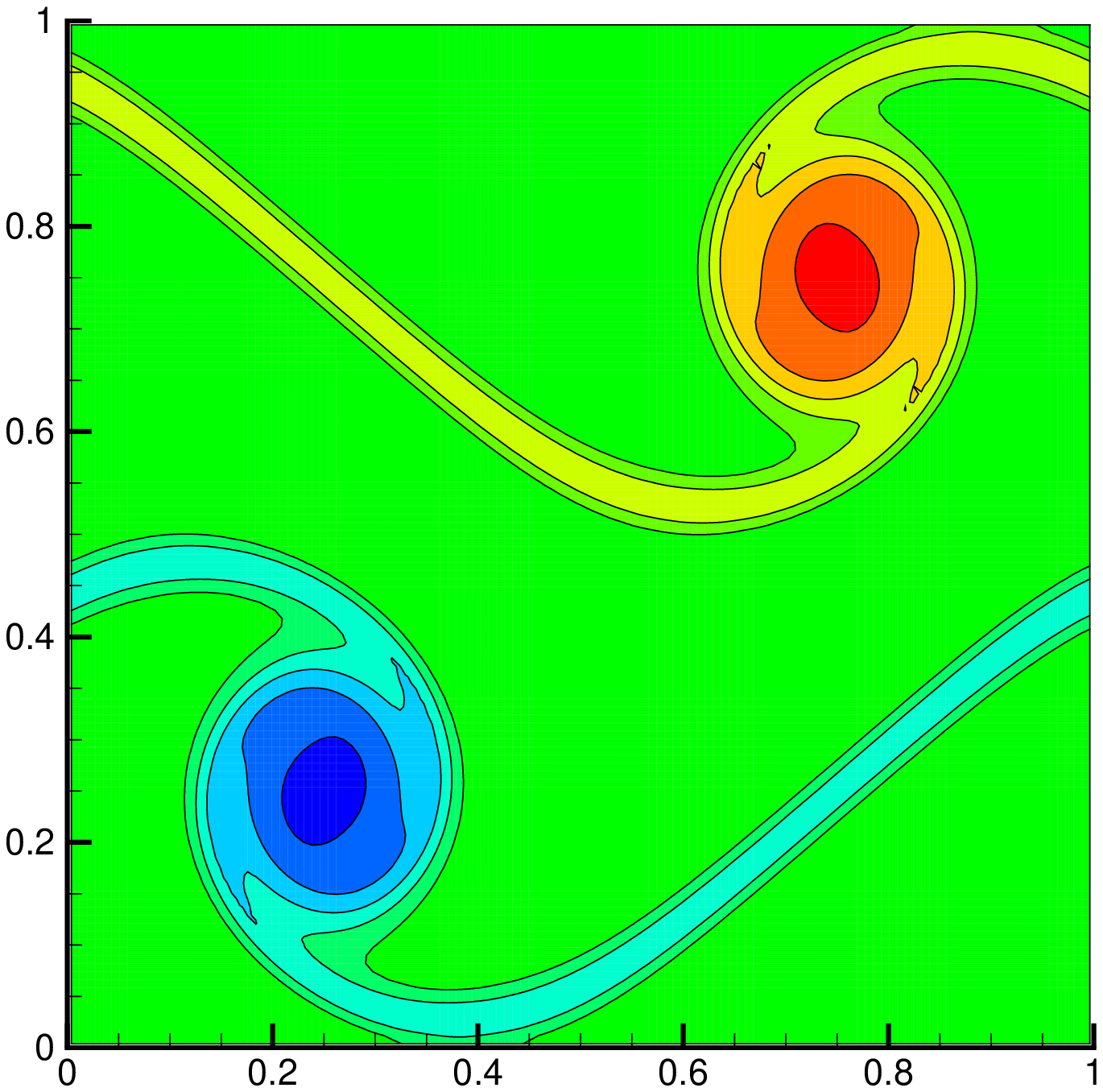}
\includegraphics[scale=0.25]{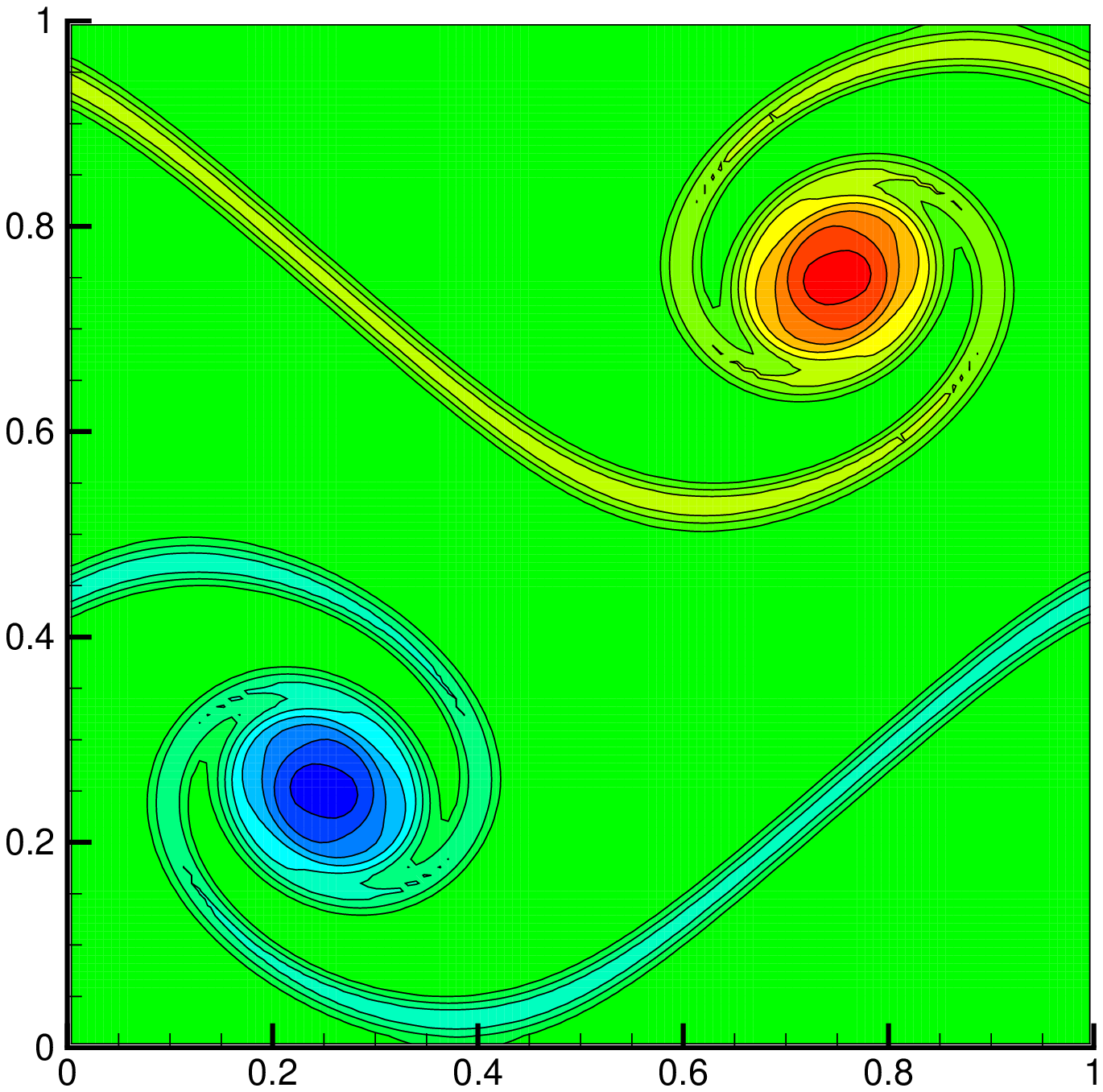}
\caption{Vorticity contours of the double shear flow at $t=1$ with $\nu = 10^{-5}$. From left to right: our implementation, the EELBM, and the bisection method.}
\label{Fig:vorticity}
\end{center}
\end{figure*}

\begin{figure*}[!ht]
\begin{center}
\includegraphics[scale=0.25]{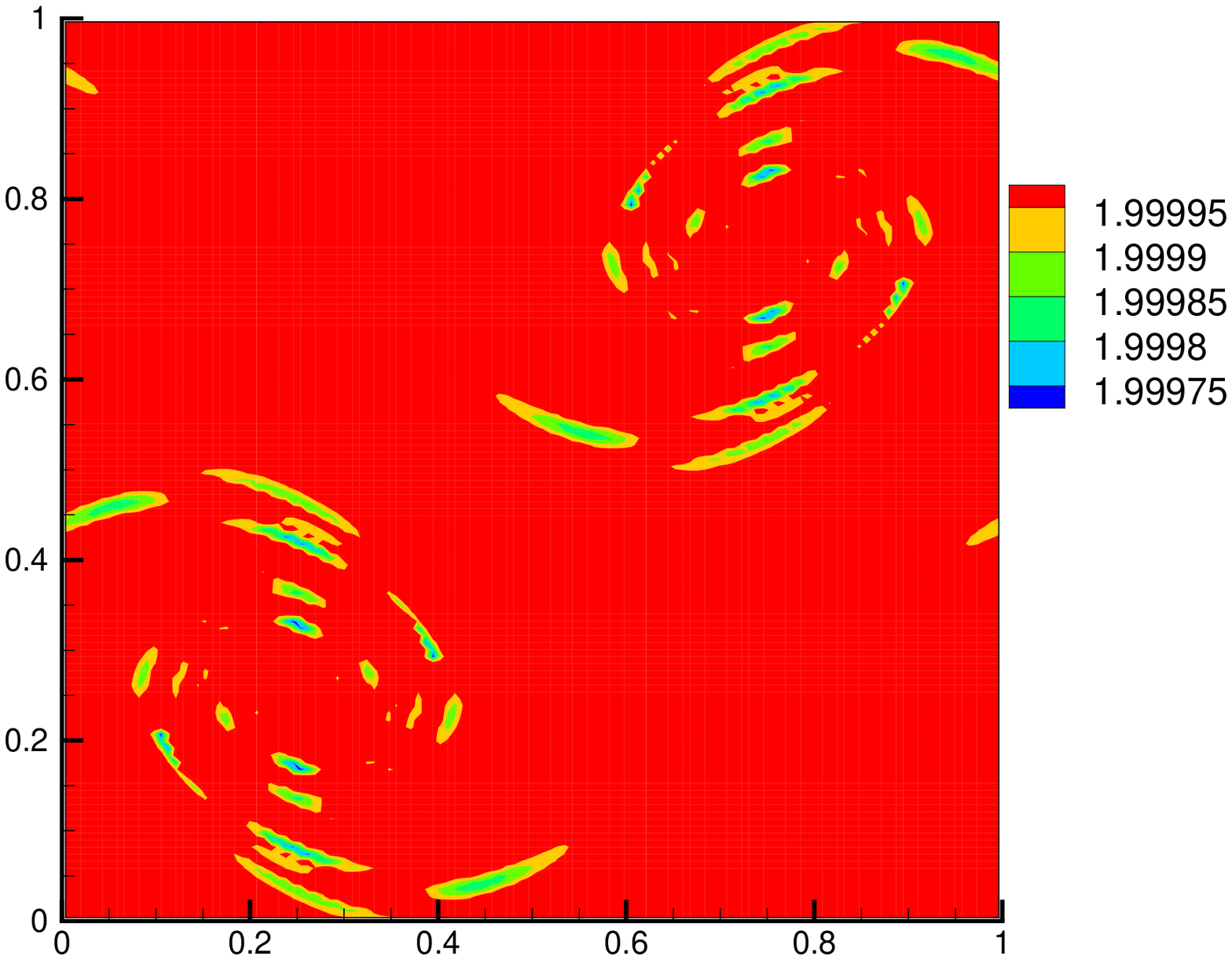}
\includegraphics[scale=0.25]{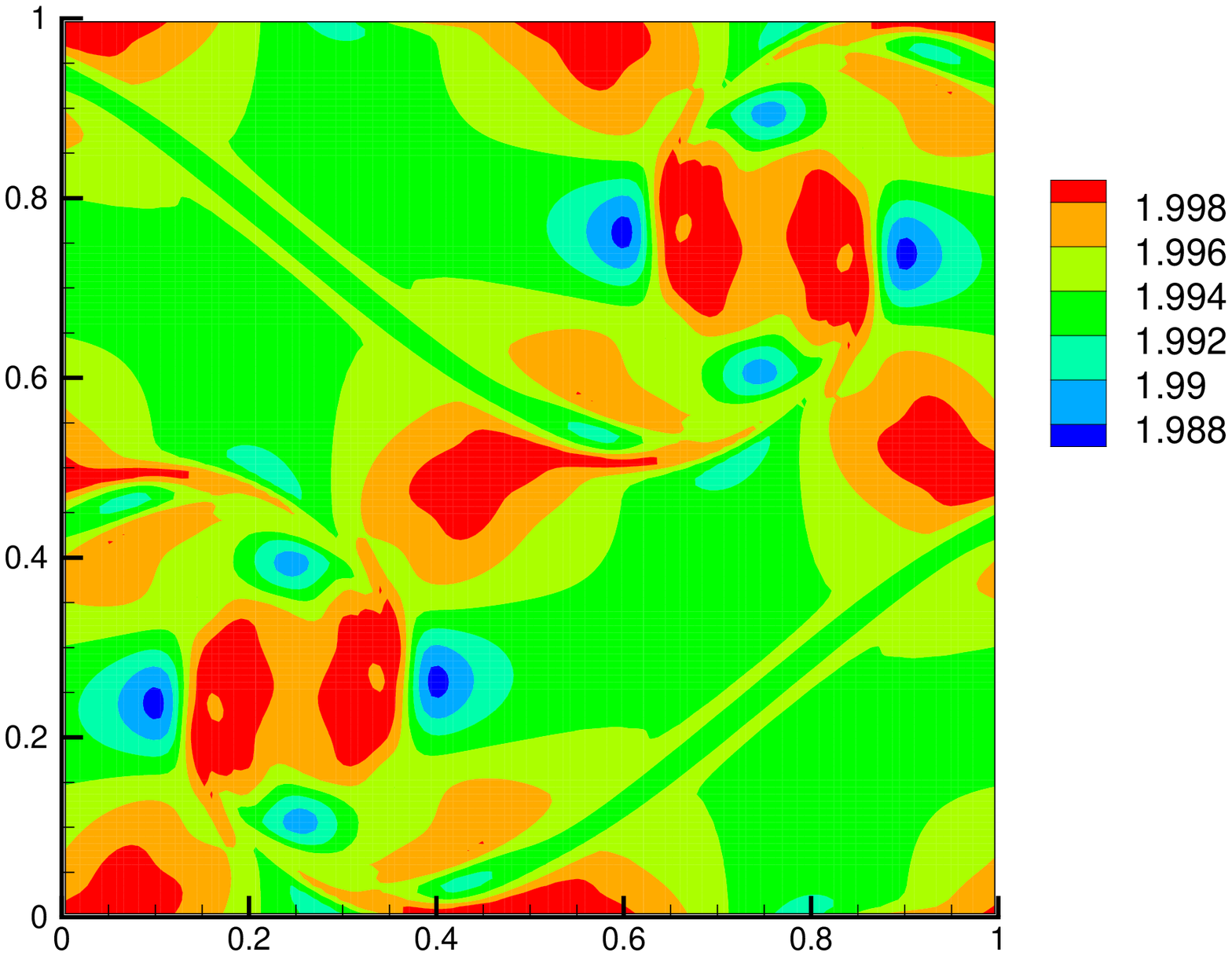}
\includegraphics[scale=0.25]{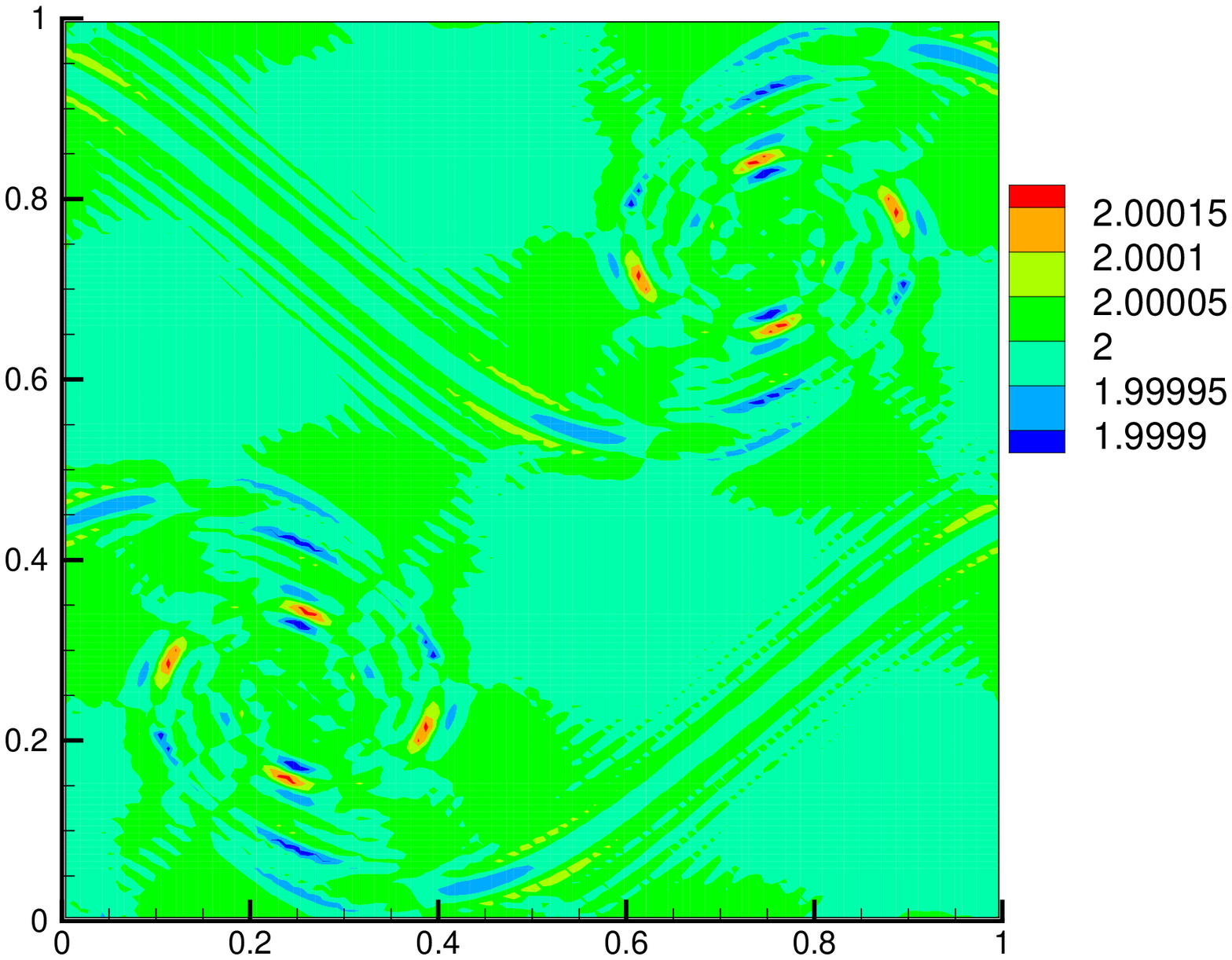}
\caption{Distributions of $\alpha$ at $t=1$ with $\nu = 10^{-5}$. From left to right: our implementation, the EELBM and the bisection method.}
\label{Fig:alpha}
\end{center}
\end{figure*}

To show the effectiveness and efficiency of our implementation, we solve this problem with three approaches: our implementation, the first-order EELBM \cite{AKTA2017}, and the bisection method used in \cite{{BGL2007}} for the equation \eqref{4}. The stop criterion for the bisection method is $-10^{-13} \leq H(f^{*}(\alpha)) - H(f) \leq 0$. Here we use the D2Q9 lattice, for which the equilibrium minimizing the entropy \eqref{3} has an analytical expression \cite{AKO2003}, and the mesh size is taken as $M^2 = 128\times 128$. For this setup, the results produced with the three approaches are displayed in Fig.~\ref{Fig:vorticity}.
They are the contour lines of vorticity at $t=1$ ($t=TU/M$, where $T=3200$ is the number of time steps and $M=128$ is the mesh size). It can be seen that, except for the EELBM, the other two methods yield almost the same shape of vortex that is consistent with those in \cite{MB1997,KBC2014}.
Furthermore, we also plot the distributions of $\alpha$ in Fig.~\ref{Fig:alpha} for the three implementations above. It clearly shows that our $\alpha$ is smaller than 2 near the vortexes and close to 2 elsewhere.
These demonstrate the efficiency of our implementation.

To have a closer look at our implementation, we also count the number, by $P[H(f^{*}(2)) \leq H(f)]$,  of the grid points where $H(f^{*}(2)) \leq H(f)$ at each time step. The proportion $P[H(f^{*}(2)) \leq H(f)]/M^2$ as a function of time $t$ is plotted in Fig.~\ref{Fig:ratio}. We see that the proportion is around $0.5$ for most of the time. Namely, at each time step Eq.~\eqref{4} needs to be solved only for about half of the total grid point number. Therefore, the introduction of $\alpha=2$ in \eqref{8} enhances the efficiency significantly.

In conclusion, we have presented a simple and uniform relaxation-rate formula for the ELBM.
The formula solves a long-standing critical problem in kinetic theory approach: it not only guarantees the nonnegativity of the distributions and the discrete H-theorem, but also gives full consideration to the consistency with hydrodynamics.
We demonstrate that with our new implementation based on the relaxation-rate formula the algebraic equation only needs to be solved for about half of the grid points, where the flow fields change drastically is effectively marked, and the computational overhead of the ELBM is greatly reduced. Thus the novel relaxation-rate formula makes a significant step for the development of the ELBM and now the method can be efficiently used for a broad range of hydrodynamics applications including turbulence flows.

Finally, we point out that Formula \eqref{9} can be improved with the following iteration (a modified secant algorithm)
\begin{align*}
\alpha^{(k+1)} = \alpha_* + \frac{ H( f^{*}(\alpha_*)) - H(f) }{ H( f^{*}(\alpha_*)) - H(f^{*}(\alpha^{(k)}))  } ( \alpha^{(k)} - \alpha_*  )
\end{align*}
for $k=1,2,3,\cdots$. In the Supplementary Material it is proved that this iteration converges unconditionally to the solution of Eq.~\eqref{4} and $H(f^{*}(\alpha^{(k)})) \leq H(f)$ for all $k$.

\end{document}